\documentclass[12pt,a4paper]{article}

\usepackage{feynmp,amsbsy,latexsym,cite}
\usepackage[dvips]{pstcol}
\usepackage{amsfonts}
\newcommand{\bc}{\begin{center}}
\newcommand{\ec}{\end{center}}
\newcommand{\no}{\noindent}

\newcommand{\Ha}{{\cal H}_{\mathrm{int}}}

\newcommand{\dd}{{d^{\dag}}}

\newcommand{\ecd}{{\cdot}}

\newcommand{\xb}{{\boldsymbol x}}
\newcommand{\yb}{{\boldsymbol y}}

\newcommand{\pb}{{\boldsymbol p}}
\newcommand{\qb}{{\boldsymbol q}}
\newcommand{\kb}{{\boldsymbol k}}
\newcommand{\vb}{{\boldsymbol v}}
\newcommand{\rb}{{\boldsymbol r}}
\newcommand{\wb}{{\boldsymbol w}}
\newcommand{\ub}{{\boldsymbol u}}

\newcommand{\intx}{\int\!d^3x\;}
\newcommand{\inty}{\int\!d^3y\;}
\newcommand{\intuv}{\int\!d^3u \, d^3v\;}
\newcommand{\intrw}{\int\!d^3r \,d^3w\;}
\newcommand{\intp}{\int\!\frac{d^3p}{(2\pi)^3}\;}
\newcommand{\intq}{\int\!\frac{d^3q}{(2\pi)^3}\;}
\newcommand{\intk}{\int\!\frac{d^3k}{(2\pi)^3}\;}

\begin{document}
\begin{center}
{\Large\bf Asymptotic Dynamics in Quantum Field
Theory}\footnote{Talk presented by R.~Horan}\\
\medskip
{\large \bf When does the coupling switch off?}\\
\medskip
\end{center}
\bigskip
\begin{center}
\textsc{Robin Horan\footnote{email: rhoran@plymouth.ac.uk}, Martin
Lavelle\footnote{email: mlavelle@plymouth.ac.uk} and David
McMullan}\footnote{email: dmcmullan@plymouth.ac.uk}\\ [5truemm]
\textit{School of Mathematics and Statistics\\ The University of
Plymouth\\ Plymouth, PL4 8AA\\ UK}
\end{center}

\bigskip\bigskip\bigskip
\begin{quote}
\textbf{Abstract:} We discuss the approach to asymptotic dynamics
due to Kulish and Faddeev. We show that there are problems in
applying this method to theories with four point interactions. The
source of the difficulties is identified and a more general method
is constructed. This is then applied to various theories including
some where the coupling does switch off at large times and some
where it does not.
\end{quote}

\bigskip

\no \textbf{Introduction:}

\medskip

\noindent In most textbook descriptions of scattering it is
tacitly assumed that the coupling \lq switches off\rq\ at large
times. This is sometimes dignified with the name \lq adiabatic
approximation\rq.
 The {LSZ} formalism, which forms the basis for the
description of scattering processes in quantum field theory, thus
assumes that at large times, particles obey a free dynamics. When
this is not justified, we have to pay the price of infra-red
divergences. However, there is a body of work concerned with
investigating the asymptotic interaction Hamiltonian and whether
or not the above assumption is justified. In this talk we will
describe a general method for the study of the asymptotic
interaction and apply it to some concrete examples.

The physical importance of this problem is clear.
 It is well known that in gauge theories, such as QED
and QCD, this assumption is incorrect and that the physics is
characterised by long range interactions. In QED, for example, the
masslessness of the photon means that the potential between static
charges falls off only as $1/r$. It is well known that
perturbative calculations of the S-matrix for the unbroken gauge
theories of the Standard Model have infra-red divergences.

It is important that we have a precise general description of the
scattering process in quantum field theory so that when looking at
theories such as QCD and other theories of the Standard Model,
where our intuition is not so developed and greater reliance must
be placed on the mathematics, we may have confidence in the
correctness of our deductions. In order to gain this confidence,
our methodology must first be tested against those theories that
are well understood. Only then can we try to understand the
asymptotic dynamics which determines hadronic spectra.

In~\cite{Dollard:1964},  working in the context of
non-relativistic Coulombic scattering, Dollard showed how the
asymptotic dynamics could be described by replacing the
interaction  Hamiltonian for the theory with a different
Hamiltonian, which we can think of as being the {\it asymptotic}
Hamiltonian.  Subsequently this approach was studied in a
relativistic setting, (for QED) by Kulish and
Faddeev~\cite{kulish:1970} and this approach has been utilised by
various other authors in QED and QCD, with varying degrees of
success, for references see~\cite{Horan:1999ba}

We shall start with the best approach to asymptotic dynamics
previously available, that of Kulish and
Faddeev~\cite{kulish:1970}. We will give a brief outline of the
method described in their article. We will then show that there
are difficulties which arise when trying to apply this method to
theories other than QED, which is the theory investigated in their
article.

After this we shall describe our general approach to asymptotic
dynamics~\cite{Horan:1999ba} which is based upon that of Kulish
and Faddeev but with two principal modifications: we work at the
level of matrix elements rather than operators and we note that
the asymptotic interaction can only be expected to vanish if the
particles are at asymptotic times a long way away from each other.

The initial description of our approach to the study of asymptotic
dynamics will be in the context of $\phi^4$ interaction theory,
after which we shall look at some applications of our method, in
particular to the three and four point interactions of {\it
scalar} QED. Finally we shall draw some conclusions and discuss
some possible further avenues of investigation.

\bigskip

\noindent{\bf A Brief Outline of the Kulish-Faddeev Method}

\medskip

\noindent Our starting point is the usual interaction Hamiltonian
for QED
\begin{equation}
  \Ha(t)=-e\intx A_\mu(t,\xb)J^\mu(t,\xb)\,.
\end{equation}
Working in the interaction picture we may insert the plane wave
expansions
\begin{equation}
\psi(x)=\intp\frac1{\sqrt{2E_{\smash{p}}}}
\{b(\pb,s)u^s(p)e^{-ip\ecd x} +  \dd(\pb,s)v^s(p)e^{ip\ecd x}\}\,,
\end{equation}
 and
\begin{equation}
A_\mu(x)=\intk\frac1{2\omega_{\smash{k}}}\{ a_\mu(\kb)e^{-ik\ecd
x}+ a_\mu^\dag(\kb)e^{ik\ecd x} \}\,,
\end{equation}
\noindent where $E_p=\sqrt{|\pb|^2+m^2}$ and {$\omega_k=|\kb|$ are
the usual energy terms.

Following Kulish and Faddeev we now substitute the plane wave
expansions of $\psi$ and $A_\mu$ into the interaction Hamiltonian
and integrate out the spatial variable $\xb$, obtaining a momentum
$\delta$-function. After integrating out this  $\delta$-function,
the resulting integrals may be conveniently grouped according to
the frequency components. Each integral will have a time
dependence of the form  $e^{i\Phi t}$, where $\Phi$ is made up of
sums and differences of the energy terms.

The substance of the  Kulish-Faddeev method is then contained in
the following claims~\cite{kulish:1970}:
\begin{itemize}
  \item  Those terms in which $\Phi$
 never vanish, such as the term of the form
$$\Phi=E_{p+k}+E_p+\omega_k\,,$$ decrease sufficiently rapidly as
$t\rightarrow \pm\infty$ and the coupling may here be set to zero.
\item  Those terms which can vanish for some values
of the momenta, e.g. the term $$\Phi=E_{p+k}-E_p-\omega_k\,,$$
which is zero for vanishing $\kb$, determine the desired behaviour
of
 $\Ha$. If there are such terms then we cannot invoke the \lq
 adiabatic approximation\rq.
\end{itemize}
It is of course highly gratifying here to note that this term only
vanishes for soft photon momenta -- which one immediately
identifies as the source of the infra-red divergences of the LSZ
scheme in QED. In a minor extension of the above calculation, one
can show that giving the photon a small mass would, according to
Kulish and Faddeev let us put the coupling asymptotically to zero
and indeed that step, as is well known, regulates infra-red
divergences.

Before one can apply this to other gauge theories, and especially
QCD, it seems reasonable to test these claims by applying the
method  to some toy models. The most straightforward examples of
quantum field theories are massive scalar theories. They are
text-book examples with no infra-red problems. Indeed for the
simple case of $\phi^3$ theory, which as far as the above argument
is concerned is similar to QED with a photon mass, the method
would imply that the coupling may be asymptotically set to zero.
This happy situation, though, breaks down when we go to a four
point interaction.

\bigskip

\noindent{\bf Application of the Kulish-Faddeev Method to $\phi^4$
Theory}

\medskip

\noindent In this theory the interaction Hamiltonian is
\begin{equation}
\Ha=\frac{\lambda}{4!}\intx :\,\phi ^4 (\xb\,,t)\,,
\end{equation}
\noindent where the free field expansion for  $\phi$ is just
\begin{equation}
\phi(\xb )= \intk \frac{1}{2E_k}\left(a(\kb)e^{-ik\cdot
x}+a^\dag(\kb)e^{ik\cdot x}\right) \,.
\end{equation}
Following the above procedure in this case, results in $12$ terms,
one of which has the form
\begin{equation}
\Phi=E_p+E_q-E_k-E_{p+q-k}\,.
\end{equation}
The Kulish-Faddeev method then implies that problems may occur if
$\Phi=0$  has any solutions. However, it is easy to
show~\cite{Horan:1999ba} that this equation has a {\it  continuum}
of solutions!

From this we conclude that  the vanishing of $\Phi$ tells us
nothing about the asymptotic behaviour of a scattering process.
This should not be surprising when one recalls that the
Kulish-Faddeev method is concerned with the {\it operator}
convergence of the interaction Hamiltonian. It is well known (see
p.~126 of \cite{thirring:1979} or Sect.~5-1-2
of~\cite{itzykson:1980}) that if the interaction Hamiltonian
converges to zero in norm, then {\it all} the fields are free
fields. We should therefore work at the level of  {\it matrix
elements}, and this we now proceed to do.

\bigskip

\noindent{\bf General Approach to  Asymptotic Dynamics}

\medskip

\noindent We consider the matrix element of the $\phi^4$
scattering process which involves two incoming and two outgoing
wave packets. The incoming and outgoing wave packets are given by
the following expressions:
 \begin{equation}  \Psi
_{\mathrm{IN}}=\intrw f(\rb)g(\wb)a^\dag (\rb)a^\dag
(\wb)|0\!\!>\,,
\end{equation}  and
\begin{equation}  \Psi _{\mathrm{OUT}}=\intuv
h(\ub)i(\vb)a^\dag (\ub)a^\dag (\vb)|0\!\!> \,,
\end{equation}
where
$f,g,h,i$ are {\it test functions} for the respective
incoming/outgoing particles.

The matrix element
$<\!\!\Psi^\dag_{\mathrm{OUT}}\,|\,\Ha(t)|\,\Psi_{\mathrm{IN}}\!\!>$
 is then proportional to

 \begin{equation}
 \int\!d^3p\, d^3q\, d^3k\, h(\kb )i(\pb +\qb -\kb)f(\pb
)g(\qb )e^{-it\Phi}\,,
\end{equation}
\medskip
where the exponent   $\Phi$ has the value
\begin{equation}
\Phi=E_p+E_q-E_k-E_{p+q-k}\,.
\end{equation}
Note that this exponent is the one which caused a problem in our
attempt to apply the Kulish-Faddeev  method to $\phi^4.$ Unlike
that case, the expression is now a genuine integral, i.e. a
c-number, so we may use elementary methods to investigate the
large time behaviour.  We shall apply the method of stationary
phase~\cite{Horan:1999ba}.

\medskip

This says that, provided there are no points in the region of
integration at which all first order partial derivatives of $\Phi$
vanish, i.e. there are no stationary points, then the integral
vanishes as $t\to\pm\infty.$

The partial derivatives of  $\Phi$ in this case are
\medskip
\begin{eqnarray}
\frac{\partial \Phi}{\partial p_i} &=&
\frac{p_i}{E_p}-\frac{p_i+q_i-k_i}{E_{p+q-k}}\,, \\
&&\nonumber\\
\frac{\partial \Phi}{\partial q_j}
&=&\frac{q_j}{E_q}-\frac{p_j+q_j-k_j}{E_{p+q-k}} \,,\\
&&\nonumber\\
 \frac{\partial \Phi}{\partial k_n} &=&
 \frac{k_n}{E_k}-\frac{p_n+q_n-k_n}{E_{p+q-k}}\,.
\end{eqnarray}

\noindent The simultaneous vanishing of these implies
\begin{equation}
\frac{\pb}{E_p}=\frac{\qb}{E_q}.
\end{equation}
Since there are no infra-red divergences in this theory, we need
to see that all the points in the region of this integration for
which this last equation holds can be excluded.

\medskip We now make the following observations:
\begin{itemize}
\item The test functions  $f,g$
for the incoming wave packet, have the arguments  $\pb,\qb$
respectively, in the expression
$<\!\!\Psi^\dag_{\mathrm{OUT}}\,|\,\Ha(t)|\,\Psi_{\mathrm{IN}}\!\!>$
\item  The expressions  $ \pb/E_p $  and
$ \qb/E_q$  represent the {\it velocities} of the  particles in
the incoming wave packet.
\end{itemize}

\noindent The desired asymptotic behaviour will thus be obtained
if the supports of $f,g$ are defined in such a way such that they
exclude the possibility that {$\pb/E_p=\qb/E_q$ }. The precise
statement for this is

\bigskip

The test functions  $f,g$ must have non-overlapping support in
velocity space.
\bigskip

This condition on the test functions, of having non-overlapping
support in velocity space, is, however,  central  to the
LSZ-formalism and the construction of the
S-Matrix~\cite{Horan:1999ba}. We therefore see that the conditions
required by LSZ fully suffice to guarantee vanishing asymptotic
dynamics for massive $\phi^4$ theory. This is in complete accord
with perturbative calculations in that theory being free of
infra-red divergences.

However, we are discounting the following situation where although
the incoming particles are separated, the outgoing particles are
clearly not free at large times

\medskip
\begin{figure}[!hbp]
\begin{center}
\includegraphics{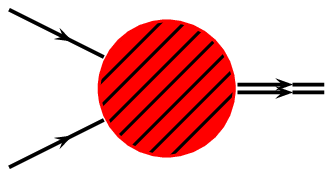}
\end{center}
%\caption{}
%        \label{}
\end{figure}
%\medskip
In this case, we would {\it not} expect to have vanishing
asymptotic dynamics  for large positive times. Such an outcome can
be discounted here, however, as we now explain.
 We have the following conservation laws which must apply:
\begin{equation}
\begin{array}{lcc}
 \mathrm{ Momentum }&: &  \pb+\qb=2\kb\\  \mathrm{ Energy}
&:&E_p+E_q=2E_k\,,
\end{array}
\end{equation}
 A simple mathematical argument~\cite{Horan:1999ba} may be used to show that
these two conditions are incompatible with the separation of the
incoming particles.
 Other matrix elements associated with this four-point interaction
may be treated in a similar fashion. This approach  shows  that
for massive ${\phi^4}$ theory we  have vanishing asymptotic
dynamics, and this is consistent with the theory having no
infra-red problems.

\bigskip

\noindent{\bf Asymptotic Dynamics  of  Scalar QED}

\medskip
\no We shall now apply this approach~\cite{Horan:1999ba} to {\it
scalar} QED. We know that  scalar QED has both three and four
point interactions.  Its perturbation theory tells us that it has
the same infra-red structure as fermionic QED -- since the
infra-red divergences can be shown to be independent of the spin
of the matter fields as long as these are massive. We would
therefore expect its four point interaction term to have trivial
asymptotic dynamics. This is reminiscent of what we have just
seen, but now two of the fields (the $A^\mu$'s) are massless.

The interaction Hamiltonian for scalar QED is then
\begin{equation}
\Ha(t) =-e\intx :J^\mu(\xb )A_\mu (\xb ):\,
\end{equation}
where the current { $J_\mu$} is given by
\begin{equation}
 J^\mu=i(\phi^\dag \partial^\mu\phi
-\partial^\mu\phi^\dag\phi)-e g^{i\mu}A_i\phi^\dag\phi\,.
\end{equation}
The plane wave expansion for { $\phi$} is
\begin{equation}
\phi(\xb )= \intp
\frac{1}{2E_p}\left(a(\pb)e^{-ip\cdot x}+b^\dag(\pb) e^{ip\cdot
x}\right) \,.
\end{equation}

We now want to consider (scalar) electron-photon scattering as
shown in the diagram.

\begin{figure}[!hbp]
\begin{center}
\includegraphics{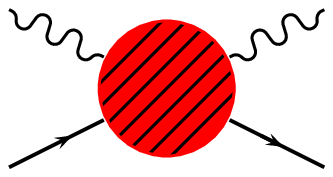}
\end{center}
%\caption{}\label{}
\end{figure}
\noindent We shall therefore examine the asymptotic dynamics of
the matrix element corresponding to the following wave packets:
\begin{eqnarray}
\Psi_{\mathrm{IN}}&=&\intrw f(\rb )b^\dag (\rb )\,c^\nu(\wb
)a^\dag_\nu(\wb )|0\!\!> \,,\\
&&\nonumber\\
\Psi_{\mathrm{OUT}}&=&\intuv g(\ub )b^\dag(\ub )\,h^\mu (\vb
)a^\dag _\mu (\vb )|0\!\!>\,.
\end{eqnarray}
\noindent  Here  $f, c^\nu, g$ and $h^\mu$ are the respective test
functions. The matrix element under consideration
$<\!\!\Psi^\dag_{\mathrm{OUT}}\,|\,\Ha(t)|\,\Psi_{\mathrm{IN}}\!\!>$}
is then proportional to
\begin{equation}  \int  d^3p\, d^3q\,
d^3k\, c^i(\pb )h_i(\qb )g(\kb )f(\kb +\qb -\pb )e^{i\Phi t} \,,
\end{equation}
where the phase function  becomes
\begin{equation}
\Phi=\omega_p +E_{k +q -p}-E_k-\omega_q\,.
\end{equation}
We now make the following deductions concerning this matrix
element: physically the incoming and outgoing charged fields must
be separated; the test functions  $c_i, h_i$ must then
automatically have disjoint supports; therefore at least one of
them, say $c_i$,  will then {\it not have zero in its support}.
With this condition the function
\begin{equation}
\Phi=\omega_p +E_{k +q -p}-E_k-\omega_q \,,
\end{equation}
now has continuous partial derivatives in both the  $\pb$ and
$\kb$ variables.

We can therefore apply the method of stationary phase to the
integral (but now with respect to the  $\pb$ and  $\kb$ variables
only).

The partial derivatives of $\Phi$ are
\begin{eqnarray}
\frac{\partial \Phi}{\partial
p_i}&=&\frac{p_i}{\omega_p}-\frac{k_i+q_i-p_i}{E_{k+q-p}} \,,\\
&&\nonumber\\
\frac{\partial\Phi}{\partial
k_j}&=&\frac{k_j+q_j-p_j}{E_{k+q-p}}-\frac{k_j}{E_k}\,.\nonumber
\end{eqnarray}

\noindent If there is a point at which  {\it all} of these vanish
then we would require:
\begin{equation}
 \frac{\pb}{\omega_p}=\frac{\kb}{E_k}
 \,.
\end{equation}
This is, however, {\it impossible}, since the first vector has
unit length but the second does not, so the integral must indeed
vanish as $t\to\pm\infty$. (Other matrix elements associated with
the four point interaction term may, we note, be treated in
exactly the same manner.)

This shows us that in scalar QED the four point interaction does
not survive asymptotically. As mentioned above, this is completely
consistent with our knowledge of perturbation theory for scalar
QED since loops involving this vertex do not generate infra-red
divergences.

\medskip

\goodbreak
\noindent{\bf Scalar QED: the Infra-Red Approximation}

\medskip \noindent Having seen which vertices do {\it not} survive
asymptotically, it is time to turn to those that do. Infra-red
divergences in perturbative calculations indicate when the
coupling does not switch off.  In such cases, we would expect this
to arise from our formalism.

Since perturbative calculations tell us that the infra-red problem
in the scalar theory is the same as that for {\it fermionic} QED,
we would expect the asymptotic properties to be the same. We now
look at the asymptotic behaviour of the scattering process
associated with the emission of a soft photon.

\begin{figure}[!hbp]
\begin{center}
\includegraphics{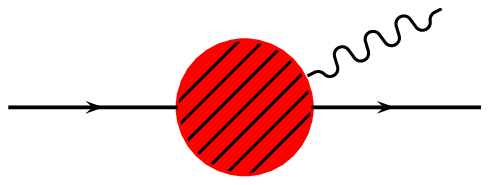}
\end{center}
%\caption{}\label{}
\end{figure}
%\medskip
The wave packets for this scattering process are taken to be
\begin{eqnarray}
\Psi_{\mathrm{IN}}&=&\inty f(\yb )b^\dag(\yb) |\,0\!\!> \,,\\
 \Psi_{\mathrm{OUT}}&=&\intuv g(\ub
)b^\dag(\ub)\,h^\mu(\vb )a^\dag_\mu(\vb)|\,0\!\!>\,.
\end{eqnarray}
The associated matrix element is then
$<\!\!\Psi^\dag_{\mathrm{OUT}}\,|\,\Ha(t)|\,\Psi_{\mathrm{IN}}\!\!>$
and when the appropriate calculations are made, it will be found
that this expression is
\begin{equation}   -e \int d^3 q\, d^3k
 f(\qb +\kb )g(\qb )q^\mu h_\mu(\kb ) e^{-i\Phi t}
 \,,
\end{equation}
where $\Phi = E_{q+k}-E_q-\omega_k$.

Rather than trying to see if this vanishes, we will show that the
difference between this interaction and another, simpler one does.
Consider then the interaction in which, instead of
$J^\mu_\mathrm{as}(\xb )$, we take the {\it asymptotic} current
given by the expression
\begin{equation}
i\intq \left( \frac{1}{2E_q}\right)^2b^\dag
(\qb)b(\qb)q^\mu \delta^3 ( \xb -\frac{\qb }{E_q}t)
\,,
\end{equation}
This current has a much simpler dynamics and a simple classical
interpretation~\cite{kulish:1970}.

The Hamiltonian associated with this asymptotic current is
\begin{equation}
 \Ha ^{\mathrm{as}} =-e\intx
J^\mu_{as}(\xb )\,A_\mu (\xb ) \,,
\end{equation}
and  the corresponding matrix element
\begin{equation}
<\!\! \Psi_{\mathrm{OUT}}^\dag\,|\Ha^{
\mathrm{as}}(t)|\,\Psi_{\mathrm{IN}}\!\!>
\,,
\end{equation}
is given by
\begin{equation}
-e \int d^3 q \,d^3k f(\qb )g(\qb )q^\mu h_\mu(\kb ) e^{-i\Phi 't}
\,,
\end{equation}
where  $\Phi '=\qb\cdot\kb/E_q-\omega_k$.

We have proven the following~\cite{Horan:1999ba}:

\medskip

\noindent{\bf  Theorem:}

\medskip
\noindent The matrix element
\begin{equation}
<\!\!\Psi^\dag_{\mathrm{OUT}}\,|\,\Ha(t)-\Ha^{
\mathrm{as}}(t)\,|\,\Psi_{\mathrm{IN}}\!\!>
\,,
\end{equation}
vanishes as {$t\to\pm\infty$}.

\bigskip
\no The interaction Hamiltonian therefore has, asymptotically, a
simpler form. This theorem, together with the previous result,
tells us that in scalar QED only the three point interaction has
non-vanishing asymptotic dynamics and its weak limit is exactly
the same as that for spinorial QED, which is indeed what
perturbation theory tells us.

\bigskip
\goodbreak

\noindent{ \bf Conclusions}

\medskip

\noindent The Kulish-Faddeev approach~\cite{kulish:1970} to the
study of asymptotic dynamics can be made~\cite{Horan:1999ba}
precise and applicable to theories with four point interactions
but we  must work at the level of { matrix} elements with
appropriate non-overlap between incoming and outgoing states. This
is exactly consistent with the LSZ formalism, where the separation
is a precondition of the theory.

We have seen that there is no need to make the \lq naive adiabatic
approximation\rq. It is possible to determine, from the formalism,
if the interaction Hamiltonian switches off. This was argued by
Kulish and Faddeev~\cite{kulish:1970} for QED, where physical
intuition tells us that  we should \textit{not} expect to be able
to  \lq switch off\rq\ the coupling in the asymptotic region. This
is because the Coulomb interaction of QED in $d=4$ has a slow
$1/r$ fall-off. However, the Kulish-Faddeev argument for when the
asymptotic interaction Hamiltonian vanishes is flawed: in
particular we have seen that it predicts that we may not switch
off the coupling in massive $\phi^4$ theory! However, we know that
both the S-matrix and the on-shell Green's functions of that
theory are well defined.

The source of this problem is, we have shown, that one should not
demand such strong restrictions on the operators. Rather we need
to consider matrix elements. A second lacuna in the Kulish-Faddeev
approach is that they do not require,  at large times,  the
separation between particles to become large.

We were able to show that with these requirements, we could
explain the asymptotic behaviour of several theories (abelian
gauge theories and massive scalar theories).

In those theories where the asymptotic interaction Hamiltonian
does not vanish, the form of the asymptotic limit is in general
hard to obtain. For QED we employed the Ansatz which was given by
Kulish and Faddeev and were able to show rigorously, using our
approach, that the weak limit form of their result is correct.

 I recall that the non-vanishing of the coupling at large
times means that gauge transformations remain non-trivial. The
physical fields are therefore much richer than the basic
Lagrangian fields. It has been shown in D.~McMullan's talk that
using the correct physical fields the coupling does effectively
switch off. It was then shown in M.~Lavelle's talk that the use of
these fields removes infra-red divergences in QED.

 Two natural
theories where this method ought to be applied are massless QED
and QCD. The former is known to have a richer asymptotic dynamics
which, unlike massive QED, is spin dependent. In particular it has
collinear divergences which are also a prominent feature of QCD.
The Kulish-Faddeev approach has been applied to massless
QED~\cite{havemann:1985,Horan:1998im} and the resulting Ansatz for
the asymptotic interaction Hamiltonian can be inserted into our
method. For QCD there is still much more work to be done, but,
since this asymptotic dynamics is responsible for holding the
hadrons together as they approach particle detectors, it is clear
that finding its exact form would be of huge importance.

\bigskip

\noindent \textbf{Acknowledgements:} I thank the Royal Society for
a travel grant to attend this workshop and the local organisers
for their hospitality.

%\bibliographystyle{h-physrev}
%\bibliographystyle{plain}
%\bibliography{litbank1}
%\begin{thebibliography}{10}

\end{document}